\documentclass[onecolumn,showpacs,preprintnumbers,amsmath,amssymb]{revtex4}
  \usepackage{graphicx}
  \usepackage{dcolumn}
  \usepackage{bm}
  \usepackage{epsfig}
  \begin{document}
  \title{Heisenberg's universal $\ln^2s$ increase of total cross sections}

  \author{H. G. Dosch}
  \affiliation{Institut f\"ur Theoretische Physik, Universit\"at Heidelberg
  Philosophenweg 16, D-69120 Heidelberg, Germany}

  \author{P. Gauron},  
  \author{B. Nicolescu}, 
  \affiliation{LPNHE~\footnote{Unit\'e de Recherche des Universit\'es Paris 6 et Paris
  7,  
      Associ\'e au CNRS} - Theory Group, Universit\'e Pierre et Marie Curie, Tour
  12 E3, 4 
       Place Jussieu, 75252 Paris Cedex 05, 
  France}

  \date{\today}

  \begin{abstract}
  The $\ln^2s$ behaviour of total cross sections, first obtained by Heisenberg 50
  years ago, receives now increased interest both on phenomenological and theoretical
  levels. 
  In this paper we present a modification of the Heisenberg's model in connection with
  the presence of glueballs and we show that it leads to a realistic description of
  all existing hadron total cross-sections data.
  \end{abstract}      
  \pacs{13.85.Lg 11.55.Jy 12.90.+b}
  \maketitle

  \section{Introduction}
  In a remarkable paper of 1952, Heisenberg investigated production of mesons as a
  problem of shock waves \cite{hei52}. 
  One of his results was that the total cross section increases like the square of the
  logarithm of the centre-of-mass energy. 
  It is noteworthy that this result coincides with very recent calculations based on
  AdS/CFT dual string-gravity theory \cite{gid}
  or on the Colour Glass Condensate Approach \cite{fiim02} and, of course, saturates
  the Froissart-Martin bound \cite{fro61}.
  In contradistinction to the latter case however the coefficient of the $\ln^2$
  term is an estimate at finite energies and not an asymptotic bound as the one
  obtained by Lukaszuk and Martin \cite{luk67}.

  We show in this note that by modifications of the original model of Heisenberg
  motivated by the enormous progress of knowledge in the 50 years that passed thence,
  the model yields some general and even some quantitative results which describe the
  data very well.

  Our article is organized as follows : In section II we shortly discuss the original
  model of Heisenberg, in section III we modify it and compare it  with the data and
  in section IV we discuss the results, their merits and their shortcomings and we
  present our conclusions.

  \section{The Heisenberg model for the total cross-section}
  The considerations of Heisenberg concerning the total cross section are essentially
  geometrical ones, but the crucial ingredient is that the energy density and not the
  hadronic density is the essential quantity to be taken into account.
  The major part of Heisenberg's paper is related to dynamical questions of meson
  production and is for our present investigation only of interest as  dynamical
  background.

  Proton-proton collisions are considered in the centre-of-mass system and the energy
  $\sqrt{s}$  is supposed to be high enough that Lorentz contraction allows to view
  the nucleons as discs (see Fig. 1).
 
 \begin{figure}
  \begin{center}
  \begin{minipage}{3cm}
  \epsfxsize3cm
  \epsffile{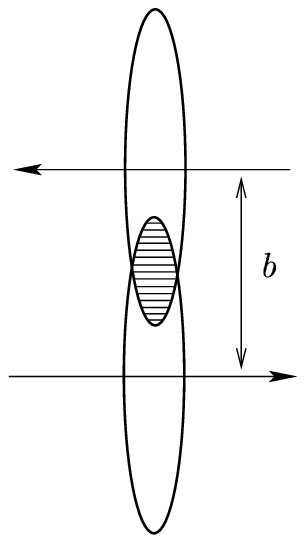}
  \end{minipage}
  \end{center}
  \caption{Scattering of two Lorentz contracted hadrons in the centre of mass
  system; the interaction region is shaded.}
  \end{figure}
  
  Interaction takes place only in the overlap region (shaded area in figure 1) and the
  crucial assumption is made that a reaction can only occur if the energy density is
  high enough in order to create at least a meson pair.

  Let $\gamma\sqrt{s}/V$ be the energy per unit volume disposable for meson
  production, where $\gamma$ is some positive constant smaller than 1.
  Accordingly to the assumption stated above, a reaction can only take place if
  $\gamma\sqrt{s}$ is sufficient to create two mesons. Il we denote the energy of the
  two mesons by $k_{0}$ we thus have the condition :
  \begin{equation}
  \label{1 }
  \gamma\cdot\sqrt{s}\ge k_{0}.
  \end{equation}
  Heisenberg took $k_{0}$ as the average energy of two produced mesons. In his shock
  wave approach with a non-renormalizable meson interaction, $k_{0}$ was only
  increasing very slowly (logarithmically) with energy.
  Next Heisenberg assumed that, at least for large impact factors $b$ (see figure
  1), $\gamma$ is proportional to the overlap of the meson clouds, that is 
  \begin{equation} \label{2 }
  \gamma=\alpha\cdot e^{-m\cdot b}
  \end{equation}
  where $\alpha$ is some constant smaller than 1 and $m$ is the mass of the mesons
  forming the cloud. From
  \begin{equation}
  \label{3 }
  \gamma\ge \gamma_{min}=\frac{k_{0}}{\sqrt{s}},
  \end{equation}
  we deduce the maximal impact parameter for which interaction takes place
  \begin{equation}
  \label{4 }
  b_{max}=\frac{-1}{m}\  \ln\ \frac{k_{0}}{\sqrt{s}\alpha}
  \end{equation}
  and therefore 
  \begin{equation}
  \label{5 }
  \sigma=2\pi\ \int_{0}^{b_{max}}b\ db=\frac{\pi}{m^{2}}\ \ln^2\
  \frac{\sqrt{s}\alpha}{k_{0}}
  \end{equation}
  which, apart from the factor $\alpha$,  is the result obtained by Heisenberg
  \cite{hei52}.
  We see that implicitely the assumption has been made that if a meson production is 
  energetically possible, it will happen (black disk).
  Of course, Heisenberg was taking the pion mass for the meson mass.
  For the energy of the produced mesons he deduced, in his dynamical considerations,
  assuming interactions of maximal strength, that the energy $k_{0}$ increases only
  slowly with energy, at any rate not by a power of $s$.
  Therefore the asymptotically leading term in  the cross section is
  $(\pi/4m^2_\pi)\ln^2s$,
  the coefficient $\pi/4m_\pi^2$ being 1/4 of the Lukaszuk-Martin bound \cite{luk67}.
  The argument can be extended easily to hadron-hadron scattering in general,
  and therefore we have the result that the coefficient of the $\ln^2s$ term is
  \textit{universal} for all hadron reactions.
     
     The assumption that the interaction is of maximal strength, that
  is highly divergent in the UV region, is essential in Heisenberg's argument. Only
  in that way enough energy is dissipated from the shock-wave front into lower
  energies to ensure copious particle production and to obtain a powerlike
  decrease of $\gamma_{min}$ in Eq. (3). Also in our present understanding
  multiparticle production is essential for the increase of cross sections with
  energy. This has been sudied in great detail in perturbative QCD by Lipatov and
  coworkers~\cite{KFL78,BL78}. The increase in phase space with $\log s$ for each
  additional particle leads to a power series in $\alpha_s \log s$ and demands
  in the high energy limit for resummation. This has been done by solving the
  famous BFKL equation. In leading order this leads to a singularity in the
  complex $J$-plane at $J=1+4 \log 2 \frac{\alpha_s N_c}{\pi} \approx 1.5$ and
  hence to a tremendous increase with energy. Though next to leading order
  corrections are huge \cite{FKL97,CC98} partial resummations seem to lead to a
  stable result~\cite{CCS99} at $J\approx 1.3$. But this can be only
  applied if at least one hard scale is present, that is, in practice, only to
  $\gamma^* $ hadron or better to $\gamma^*~\gamma^*$ scattering with high photon
  virtualities. In this case the Froissart bound cannot be derived. Nevertheless
  it seems plausible that also for scattering of small objects the asymptotic
  theorems hold. I a recent paper~\cite{fiim02} it was shown that gluon
  saturation effects together with confinement lead indeed to a moderation of
  the BFKL-type behaviour to a $\log^2 s$ behaviour at high energies. Quite
  generally speaking it is very plausible that a strong increase of the cross
  section eventually violating unitarity bounds is eventually damped to satisfy
  the Froissart bound (see for instance \cite{SSP02}). The connection
  between the Heisenberg model and the considerations of \cite{fiim02} has been
  discussed in \cite{Heb02}. Here it is suggested to get a more quantitative
  connection in a U(1) gauge theory with massive gauge bosons and fermions. This
  would in some way be a return to the model of Cheng and Wu  \cite{CW70}. 

  For hadron-hadron scattering the perturbative treatment of BFKL is not
  possible, but there has been a promising development~\cite{FK99,KL00} to
  explain the increase of the hard cross section by multiple particle production
  in a way analogous to BFKL. But, in contrast to the perturbative treatment in
  QCD, the produced ``particles'' (rungs in the BFKL ladder) are not gluons but
  colourless objects, pions for instance (see Fig. 2). In this way a
  singularity in the complex $J$-plane at $J\approx 1.1$ has been
  obtained~\cite{KL00}. It is interesting to note that in the more recent
  publications~\cite{KKL00,SZ00,COS01} emphasis is given to instanton induced
  interactions as source of many-particle production. All these attempts have in
  common that the rise of energy in hadron-hadron cross sections is a genuine
  non-perturbative effect which brings them closer to Heisenbergs original idea.

  \begin{figure} 
  \begin{center} 
  \begin{minipage}{3cm} 
  \epsfxsize3cm 
  \epsffile{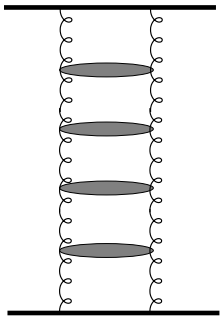} 
  \end{minipage} 
  \end{center} 
  \caption{Multiparticle production leading to an increasing cross
  section according to \cite{KL00}, the gray ellipses represent hadrons (two
  pions e.g.)  or may indicate instanton induced reactions.}  
  \end{figure}

  \section{Modifications of the model}
   There are two obvious necessary modifications of the Heisenberg
  model :
   \begin{itemize}
    \item [1)] If we want to apply it to all kind of hadrons, we have to take care of
    the different hadron sizes, since in the above treatment all sizes are equal
    to $1/m$.
    \item [2)] We have to take into account that {\em direct} pion exchange,
  though being the
   exchange with the lightest particle, is not relevant at high
  energies.
   This is due to the fact that exchanged gluons have spin 1 and pions spin
   0. Therefore already in Born approximation gluon exchange dominates
   dominates at high energies. In Regge theory this is manifested by the
   fact that intercept of the pion is much lower than that of the Pomeron. \end{itemize} 
  
   \begin{figure}
  \begin{center}
  \begin{minipage}{8cm}
  \epsfxsize8cm
  \epsffile{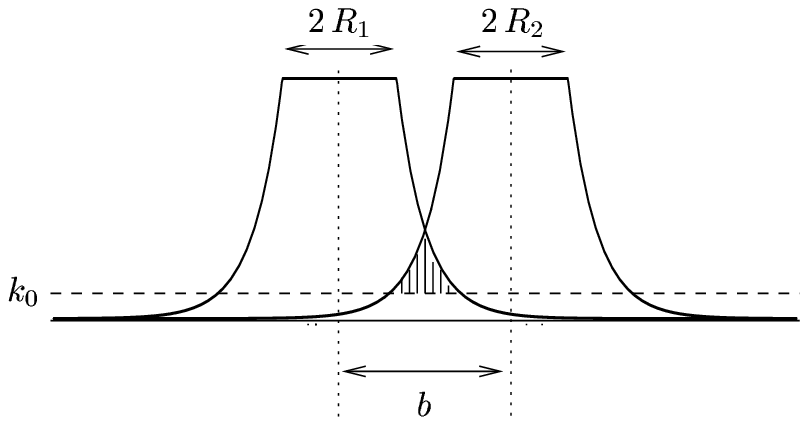}
  \end{minipage}
  \end{center}
  \caption{Interaction region (shaded) of two hadrons in the modified Heisenberg
  model of high energy scattering.} 
  \end{figure}
  
  We
  therefore modify the model in two respects : \begin{itemize}   \item [1)] We
  make a rough approximation for the overlap (see Fig. 3) : \begin{equation}
  \gamma=\left\{
  \begin{array}{ll }
   \alpha     &   \mbox{for } b\le R_1+R_2 \\
  \\
    \alpha e^{-m(b-R_1-R_2)},    &   \mbox{for } b\ge R_1+R_2
  \end{array}
  \right.
  \end{equation}
    \item [2)] For the mass we rather insert a mass $M$ in the range of the 
    glueball mass instead of the pion mass $m$, since we believe that the high-energy
    behavior is dominated by gluon 
    exchange.
  \end{itemize}
  We then obtain for $b_{max}$
  \begin{equation}
  \label{bmax }
  b_{max}=R_1+R_2+\frac{1}{M}\ \log \frac{\sqrt{s}\alpha}{k_0},
  \end{equation}
  assuming $\sqrt{s}$ large enough that $\ln\displaystyle \frac{\sqrt{s}\alpha}{k_0}>0.$

  For the total cross section we thus obtain :
  \begin{eqnarray}
  \sigma   &=&e\frac{\pi}{M^2} \ln^2\ \frac{\sqrt{s}\alpha}{k_0}+
  2\pi(R_1+R_2)\ln \frac{\sqrt{s}\alpha}{k_0}\frac{1}{M}\nonumber \\ 
  &&~~~~~~~~~~+
   \pi(R_1+R_2)^2\nonumber \\
  &=&   \displaystyle\frac{\pi}{4M^2}\ln^2s+\displaystyle\frac{\pi}{M}\ln s
  \left\{
  (R_1+R_2)+\frac{1}{M}\ln \frac{\alpha}{k_0}
  \right\}\nonumber \\
  &&+\pi(R_1+R_2)^2+\frac{\pi}{M^2}\ln^2\frac{\alpha}{k_0}
  \nonumber \\
  && ~~~~~~~~~~~~~~~+\frac{2\pi}{M}(R_1+R_2)\ln \frac{\alpha}{k_0},
  \label{sigma} \end{eqnarray}
  where, as usual, implicite scale factors of 1  $GeV^2$ and 1 GeV,
  respectively, are obviously assumed in writing $\sigma$ as a quadratic
  form in ln s.
  We see that the leading $\ln^2s$ term is still universal, but now dominated rather
  by a glueball than by the pion mass.
  Since $R_1$ and $R_2$ are supposed to be of the size of the electromagnetic radii,
  the second term in Eq. (8) will dominate over the
  $\displaystyle\frac{\pi}{M^2}$ term except at high energies, $s\gg k_0^2/\alpha^2$.

  In order to perform a rough numerical estimate, we may insert for the glueball mass
  a value between 1.4 and 1.7 GeV, yielding
  \begin{equation}
  \label{gb }
  \frac{\pi}{4M^2}=0.11-0.16\ \mbox {mb}.
  \end{equation}
  For $R_1$ and $R_2$ we may insert the electromagnetic radii. In contrast to
  Heisenberg, we insert for $k_0$ the minimal energy of two produced particles. Since
  production seems to occur in clusters with mass around 1.3 GeV \cite{gia79}, we can
  put $k_0=2.6$ GeV. The value of $\alpha\ (0\le \alpha\le 1)$ might be process
  dependent. For very small objects ("onia") it might be very small.

  In the past, application of the Heisenberg model to the global
  analyses of the forward hadronic data were performed in \cite{ey88}, but the
  universality of the leading term was not discussed there. This
  universality was treated by Gershtein and Logunov \cite{gl84}, who made the
  assumption, as in the present paper, that the growth of the hadron-hadron
  total cross-sections is related to resonance production of glueballs.

  A very good fit for all forward data has been recently obtained by assuming a
  universal $\ln^2s$ dependence and a constant contribution dependent on the process
  \cite {ga00,cud02}.
  Apart from Reggeon-exchanges, which are of no concern here, the cross section was
  fitted \cite{cud02} to
  \begin{equation}
  \label{sigmaHH }
  \sigma^{HH}=B\ln^2\left(\frac{s}{s_0}\right)+Z^{HH},
  \end{equation}
  with $B\simeq$ 0.32 mb, $Z^{pp}\simeq$ 36 mb, $Z^{\pi p}\simeq$ 21 mb, 
  $Z^{Kp}\simeq$18 mb and $s_0\simeq$ 34 GeV$^2$.

  Such a value of $B$ would correspond to a mass $M$ of 1 GeV, a bit small for a
  glueball, but not unreasonable given the crude approximations.
  $Z^{HH}$ are in the right order of magnitude of $R^2$.
  
  The scale $s_0$ is related to the quantities $\alpha$ and $k_0$ in (\ref{2 })
  and (\ref{3 }). Using for $k_0$ the cluster mass~\cite{gia79} we obtain for the
  scattering of normal hadrons a value $\alpha \approx 0.28$, that is about a
  fourth of the energy goes into particle production. 

  The term $Z^{HH}$ in (10) has been derived here from the purely
  geometrical model (8). An energy independent term has been also
  obtained in a nonperturbative model for high energy scattering. It is based on
  an functional-integral approach to high energy scattering~\cite{Nac91}, where the
  functional integrals are evaluated in a specific model for nonperturbative
  QCD~\cite{KD91}, based on the gauge invariant gluon-gluon
  correlator~\cite{Dos87,DS88}. The hadron dependent values of Z for $pp$, $\pi p$,
  and $Kp$ scattering in this model~\cite{DFK94} are 35 mb, 23 mb and 19 mb
  respectively.

\section{Conclusions}
  We have seen that a geometrical model, where the energy density available for
  particle production is the relevant quantity, leads to the quite general result that
  the leading term in energy is of $\ln^2s$ type and universal for all hadronic
  processes.
  Such a type of energy behavior has been obtained as  the best choice testing many
  models and even the parameters found in the fit to all forward data \cite{cud02} are
  in qualitative agreement with a reasonable choice of the parameters of the present
  model.

  A consequence of the universal $\ln^2 s$ term is that at asymptotic energies
  all hadron cross sections become equal. At finite but high energies the pion
  and kaon proton cross sections are therefore expected to rise somewhat faster
  than the nucleon-nucleon cross sections. This seems indeed to be indicated by
  the data.

  A crucial modification of the original model of Heisenberg consisted in replacing
  the pion mass determining the fall off of the energy density available for
  high-energy reactions by a mass close to the expected glueball mass.
  This is necessitated by the fact that {\em direct} pion exchange does not contribute to
  high-energy scattering.
  This may explain why the Lukaszuk-Martin bound could not be essentially lowered.
  If one insists on rigor, one has to use the lightest particles to take into account
  the nearest singularities, but it is quite certain that these nearest singularities
  are irrelevant for high-energy scattering.

  We are fully aware that the model proposed here is very sketchy but we think
  interesting enough to invetigate further the energy dependence of high
  energy reactions suggested by that paper. It  leaves however many questions
  open, some  of them are quite obvious : 
 \begin{itemize}   
 \item [1)] How to generalize to differential cross sections ?  
 \item [2)] How to incorporate this s-channel picture into Regge approach ?  
 \item [3)] How to include possible effects, as an asymptotically leading $C=-1$ exchange (Odderon) \cite{luk73} ?   
 \item [4)] How to unify this approach with a treatment of DIS
  and especially how to explain the sharp rise in $1/x$ for small $x$ and large
  $Q^2$ ? 
 \end{itemize}

  \section*{Acknowledgements}  It is pleasure to thank Carlo Ewerz, Vladimir Ezhela,
  Otto Nachtmann and Werner Wetzel for interesting discussions and critical comments.
  One of us (H.G.D.) thanks Professor J.E. Augustin for the hospitality at LPNHE -
  Universit\'e Paris 6, where part of this work was done.



\begin{thebibliography}{9}
  \bibitem{hei52} W. Heisenberg, Z. Phys. 133 (1952) 65.
  \bibitem{gid} S. B. Giddins, hep-th/0203004.
  \bibitem{fiim02} E. Ferreiro, E. Iancu, K. Itakura, and L. McLerran, hep-ph/0206241.
  \bibitem{fro61} M. Froissart, Phys. Rev. 123 (1961) 1053 ; A. Martin, Nuovo Cimento
  A42 (1966) 930.
  \bibitem{luk67} L. Lukaszuk and A. Martin, Nuovo Cim. A52 (1967) 122.
  \bibitem{KFL78}
  E.~A.~Kuraev, L.~N.~Lipatov and V.~S.~Fadin,
  Sov.\ Phys.\ JETP {\bf 45} (1977) 199
  [Zh.\ Eksp.\ Teor.\ Fiz.\  {\bf 72} (1977) 377].

  \bibitem{BL78}
  I.~I.~Balitsky and L.~N.~Lipatov,
  Sov.\ J.\ Nucl.\ Phys.\  {\bf 28} (1978) 822
  [Yad.\ Fiz.\  {\bf 28} (1978) 1597].

  \bibitem{FKL97}
  V.~S.~Fadin, M.~I.~Kotsky and L.~N.~Lipatov,
  Phys.\ Lett.\ B {\bf 415} (1997) 97.

  \bibitem{CC98}
  M.~Ciafaloni and G.~Camici,
  Phys.\ Lett.\ B {\bf 430}, 349 (1998)
  [arXiv:hep-ph/9803389].

  \bibitem{CCS99}
  M.~Ciafaloni, D.~Colferai and G.~P.~Salam,
  Phys.\ Rev.\ D {\bf 60} (1999) 114036
  [arXiv:hep-ph/9905566].

  \bibitem{SSP02}
  A.~I.~Shoshi, F.~D.~Steffen and H.~J.~Pirner,
  arXiv:hep-ph/0205343.

  \bibitem{Heb02}
  A.~Hebecker, {\em High Energy Scattering and Diffraction at HERA},  contribution
  to ``Strong and Electroweak Matter'', Heidelberg 2002, to appear in the
  proceedings.
  
  \bibitem{CW70}
  H. Cheng  and T.~T.~Wu,
  Phys.\  Rev.\ Lett.\ {\bf 24} (1970) 1456.

  \bibitem{FK99} H.~Fujii and D.~Kharzeev,
  Phys. Rev. D {\bf 60} 114039 (1999)
  arXiv:hep-ph/9807383.

  \bibitem{KL00}
  D.~Kharzeev and E.~Levin,
  Nucl.\ Phys.\ B {\bf 578} (2000) 351
  [arXiv:hep-ph/9912216].

  \bibitem{KKL00}
  D.~E.~Kharzeev, Y.~V.~Kovchegov and E.~Levin,
  Nucl.\ Phys.\ A {\bf 690} (2001) 621
  [arXiv:hep-ph/0007182].

  \bibitem{SZ00}
  E.~V.~Shuryak and I.~Zahed,
  Phys.\ Rev.\ D {\bf 62} (2000) 085014
  [arXiv:hep-ph/0005152].

  \bibitem{COS01}
  G.~W.~Carter, D.~M.~Ostrovsky and E.~V.~Shuryak,
  hadron collisions,''
  Phys.\ Rev.\ D {\bf 65} (2002) 074034
  [arXiv:hep-ph/0112036].

  \bibitem{gia79} G. Giacomelli and M. Jacob, Phys. Rep. C55 (1979) 1.

  \bibitem{ey88} V. V. Ezhela and O. P. Yushchenko, Serpukhov preprints
  IFVE-87-6,
   1987 (in Russian) and IFVE-88-198, 1988 (unpublished).

  \bibitem{gl84} S. S. Gershtein and A. A. Logunov, Sov. J. Nucl. Phys. 39, 960
  (1984).
 
  \bibitem{ga00} P. Gauron and B. Nicolescu, Phys. Lett. B486 (2000) 71
  ; 
   B. Nicolescu, Nucl. Phys. B (Proc. Suppl.) 99 (2001) 47.

  \bibitem{cud02} J. R. Cudell, V. V. Ezhela, P. Gauron, K. Kang, Yu. V.
  Kuyanov, S. B. Lugovsky, B.~Nicolescu, and N. P. Tkachenko, Phys. Rev. D65
  (2002) 074024. These results are included in the 2002 Review of Particle
  Physics, K. Hagiwara et al., Phys. Rev. D66, 010001 (2002).

  \bibitem{Nac91}
  O.~Nachtmann,
  Annals Phys.\  {\bf 209} (1991) 436.

  \bibitem{KD91}
  A.~Kraemer and H.~G.~Dosch,
  Phys.\ Lett.\ B {\bf 272} (1991) 114.

  \bibitem{Dos87}
  H.~G.~Dosch,
  Phys.\ Lett.\ B {\bf 190} (1987) 177.

  \bibitem{DS88}
  H.~G.~Dosch and Y.~A.~Simonov,
  Phys.\ Lett.\ B {\bf 205} (1988) 339.

  \bibitem{DFK94}
  H.~G.~Dosch, E.~Ferreira and A.~Kramer,
  Phys.\ Lett.\ B {\bf 289} (1992) 153.
  
  \bibitem{luk73} L. Lukaszuk and B. Nicolescu, Nuovo Cimento Lett. 8 (1973) 405
  ; K. Kang and B.~Nicolescu, Phys. Rev. D11 (1975) 2461 ; P. Gauron, E. Leader,
  and B. Nicolescu, Phys. Rev. Lett. 54 (1985) 2656.  

  \end{thebibliography}
  \end{document}